\documentclass[aps,prl,reprint,10pt,superscriptaddress,showpacs]{revtex4-1}

\usepackage{amsmath} 
\usepackage{amssymb} 
\usepackage{graphicx} 
\usepackage{hyperref} 
\usepackage{subfigure}
\usepackage{color}

\begin{document}

\title{Topological superconductivity in Quantum Hall--superconductor hybrid systems}

\author{Bj\"orn Zocher}
\author{Bernd Rosenow}
\affiliation{Institut f\"ur Theoretische Physik, Universit\"at Leipzig, D-04103 Leipzig, Germany}

\date{\today}

\begin{abstract}

We develop a  scenario to engineer a topological phase with Majorana edge states  based on an integer quantum Hall (QH) system 
proximity coupled to a superconductor (SC). Due to the vortices in the SC order parameter, the SC - QH  hybrid system is described by a   Bloch problem with ten unpaired momenta, corresponding to the maxima and saddle points of the SC order parameter. 
For  external  potentials respecting the symmetry of the vortex lattice, the  states with unpaired momenta  have degeneracies such that the system always is in a trivial phase.  However, an incommensurate potential can lift  the degeneracies and drive the system into a topologically nontrivial phase. 

\end{abstract}


\maketitle

{\it Introduction}.---Non-abelian anyons are zero-energy quasiparticles whose ground-state wave function is rotated in the space of degenerate ground states under particle exchange \cite{NS2008,B2013,A2012}. Searching for non-abelian anyons \cite{CA2013,LB2012,V2013,BQ2013,MC2014} is motivated both by possible applications to fault-tolerant quantum computation \cite{NS2008} and by the pursuit of exotic fundamental physics. Recently, there has been much interest in Majorana fermions (MFs), which are a prime example for non-abelian anyons \cite{B2013,A2012}. MFs are their own antiparticles, and a pair of MFs can encode a fermionic two-level system in a nonlocal way.

The first system which was proposed to realize MFs is the fractional quantum Hall (QH) state at filling fraction $\nu=5/2$ \cite{MR1991,RG2000}. Recently, the superconducting proximity effect has been suggested as a way to effectively induce $p$-wave pairing in semiconductors with strong spin-orbit coupling \cite{SLTS2010,LS2010,OR2010}, and a series of experimental works presented first evidence for MFs in such heterostructures \cite{MZ2012,AR2012,RL2012,DY2012}. However, despite the effort which has been invested, unambiguous experimental signatures of the unconventional nature of MFs are still missing. 

Here, we develop an alternative scenario to engineer a topological superconductor (SC) whose zero-energy quasiparticles are MFs. 
In contrast to previous proposals \cite{SLTS2010,LS2010,OR2010}, we consider the strong magnetic field limit where 
 a $\nu=1$ integer QH system is proximity coupled to an $s$-wave SC via a spin sensitive barrier. Due to the spin sensitivity of the barrier, an effective $p$-wave pairing potential is induced in the spin-polarized lowest Landau level (LLL). In the presence of an external magnetic field necessary to reach the QH regime, the superconducting pairing potential has a triangular Abrikosov vortex lattice imprinted, with a flux of $h/2e$ per vortex. In the LLL, there is one electronic state per flux quantum $h/e$ threading the system, such that there are two superconducting vortices per electronic state. This lattice system can be described by a pair of commuting generalized translation operators $\mathcal{T}_x$ and $\mathcal{T}_y$ which define a Bloch problem with generalized momenta ${\bf k} = (k_x, k_y)$, despite the absence of translational invariance of the Hamiltonian for electrons in a magnetic field \cite{B1964}. 

\begin{figure}[t]
\subfigure{\includegraphics[width = .45\textwidth]{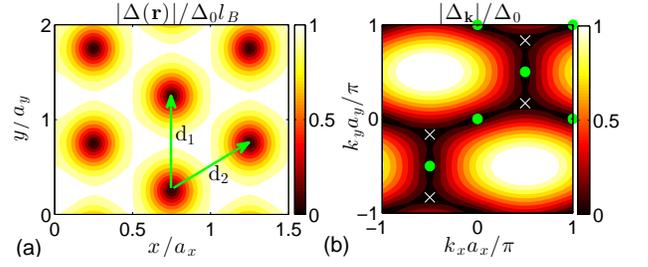}}
\caption{(a) Real space structure of the $s$-wave pairing potential $\Delta({\bf r})$  with triangular vortex lattice in the LLL approximation. The arrows denote the lattice vectors ${\bf d}_1 = a_y {\bf e}_y$ and ${\bf d}_2 = (a_x {\bf e}_x + a_y{\bf e}_y)/2$. (b) Brillouin zone of the rectangular unit cell comprising two vortices, and induced $p$-wave pairing potential $\Delta_{\bf k}\sim \partial_x \Delta({\bf r} = l_B^2 {\bf k} \times \hat{z})$. Markers denote the unpaired momenta, which correspond to the $M$ points (dots) and the $K$, $K^\prime$ points (crosses) of the hexagonal Brillouin zone of the vortex lattice.  By pairwise coupling the $M$ points, the system can by driven into a topologically nontrivial phase. }
\label{fig:BZ_2d}
\end{figure}

The superconducting pairing couples momenta ${\bf k}$ and $-{\bf k}$ with an amplitude $\Delta_{\bf k}$, similar to the situation without magnetic field. Interestingly, there are ten unpaired momenta in the Brillouin zone, since $\Delta_{\bf k}$ is represented as $\Delta_{\bf k} \sim \partial_x \Delta({\bf r} = l_B^2 {\bf k} \times \hat{z})$, with $\Delta({\bf r})$ denoting an $s$-wave superconducting pair potential with an imprinted Abrikosov vortex lattice (see Fig.~\ref{fig:BZ_2d}(a)). The unpaired momenta  arise due to the four maxima and six saddle points of $|\Delta({\bf r})|$. The zeroes of $\Delta({\bf r})$ all have a finite slope since superconducting vortices correspond to simple zeroes, and hence do not give rise to unpaired momenta. As illustrated in Fig.~\ref{fig:BZ_2d}(b), the first Brillouin zone can be decomposed into two hexagons, each associated with one vortex of the unit cell. Drawing an analogy between one such hexagon 
 and the Brillouin zone of the honeycomb lattice in graphene \cite{NG2009}, the three unpaired momenta originating from the saddle points correspond to the $M$ points, and the remaining two correspond to the $K$ and $K^\prime$ points.

\begin{figure}[t]
\includegraphics[width = .45\textwidth]{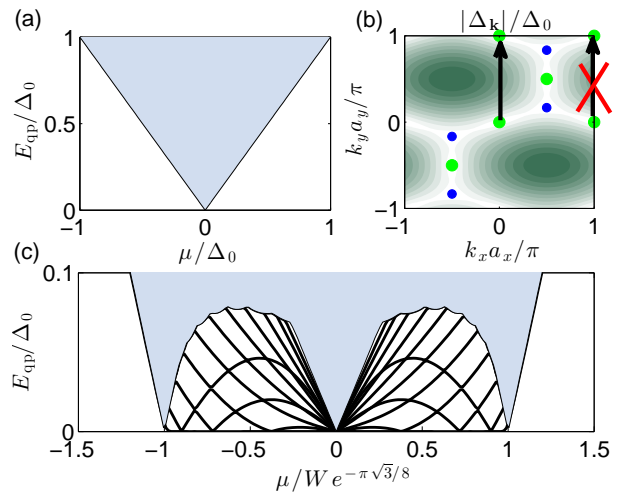}
\caption{(a) Excitation spectrum $E_{\bf k} = |\Delta_{\bf k}|$. All unpaired momenta have zero energy. (b) Coupling of unpaired momenta $(0,0)$ and $(0,\pi)$ by the external potential $V({\bf r}) = W \cos(y Q_y/2)$ lifts their degeneracy, and (c) excitation spectrum as function of chemical potential for $W = \Delta_0/2$. Shaded: Bulk spectrum for periodic BC in $x$ direction. Black lines: Majorana edge mode for open BC in $x$ direction.}
\label{fig:cosy_pot_spec}
\end{figure}

Due to the degeneracy of symmetry related unpaired momenta, the hybrid system has an even particle number parity, and hence is topologically trivial (see Fig.~\ref{fig:cosy_pot_spec}(a)). However, according to Refs.~\cite{K2001,S2009,FB2010}, it can be driven into a topologically nontrivial phase by changing the parity of the ground state, i.e.~by moving an odd number of unpaired momenta through the Fermi level. We show that this can be realized by appropriately coupling either one or three pairs of unpaired momenta ${\bf k}_{1} $ and ${\bf k}_{2}$, thus lifting the degeneracy between them (see Fig.~\ref{fig:cosy_pot_spec}(b),(c)). Experimentally, this can be achieved by engineering a periodic potential $V({\bf r})$ with Fourier components for momenta ${\bf q}$ connecting ${\bf k}_{1}$ and ${\bf k}_{2}$. Consequently, breaking the symmetry of the vortex lattice is a necessary criterion for driving the system into a topologically nontrivial superconducting phase. Since it turns out that  the momenta corresponding to the $K$ and $K^\prime$ points cannot be split pairwise, the external potential needs to couple one or three pairs of $M$ points.

Usually, only a low density of vortices in a topological $p$-wave SC is considered \cite{GS2006,KS2011,B2013a,S2013,II2002,LL2011}. Since every vortex hosts a MF, a tunnel coupling of MFs generically destabilizes the topological (Ising) phase \cite{MC2014}. However, we here consider the high vortex density limit and show that under suitably chosen external potential modulations, the Ising phase is stable even in the limit of strong coupling between the vortices.

Recently, it was proposed that ${\mathbb Z}_N$ parafermions can be realized by proximity coupling stripes of $\nu = 2/N$ fractional QH liquids to 
superconducting trenches 
 \cite{CA2013,LB2012,V2013,BQ2013,MC2014}. Based on our findings for the case $N=2$, the alternating FQH stripes and SC trenches considered in \cite{MC2014} may not only be convenient for the theoretical analysis, but actually be an important 
ingredient for the creation of exotic Ising or Fibonacci anyon phases, similarly to the potential modulation considered in this letter. 
 Thus, it is  conceivable that a spatially homogeneous FQH liquid  coupled to an SC could be tuned into an exotic anyonic phase by applying a periodic potential modulation.

{\it Model}.---We consider a two-dimensional electron gas with periodic boundary condition (BC) in $y$ direction and subject to a strong perpendicular magnetic field $B$. For fixed electron density, the magnetic field is chosen such that the LLL is (partially) occupied. We project the electron field operator onto the LLL
\begin{equation}
\hat{\psi}({\bf r}) = \sum_\kappa \hat{a}_\kappa \varphi_\kappa({\bf r})
\label{eqn:psi_LLL}
\end{equation}
with the LLL single-particle wave functions $\varphi_\kappa({\bf r})$ and fermionic annihilation operators $\hat{a}_\kappa$. We here describe the magnetic field by the vector potential ${\bf A}({\bf r}) = Bx {\bf e}_y$ in the Landau gauge. In this choice, the single-particle wave functions are 
\begin{equation}
\varphi_\kappa({\bf r}) = \frac{1}{\left(L_y l_B \sqrt{\pi} \right)^{\frac{1}{2}}} e^{-i\kappa y} e^{-\frac{1}{2}\left(\frac{x}{l_B} - \kappa l_B\right)^2}, 
\label{eqn:wf_LLL}
\end{equation}
where $L_y$ denotes the length of the system in $y$ direction, $l_B = (\hbar/eB)^{1/2}$ the magnetic length, and $\kappa=2\pi n/L_y$ with $n \in \mathbb{Z}$ denotes the momentum in $y$ direction which corresponds to the location $\langle x \rangle_\kappa = \kappa l_B^2$ in $x$ direction.

We assume that the electron gas is proximity coupled to an external singlet SC via a spin sensitive barrier which allows for spin flips, such that it effectively couples like a spinless $p$-wave SC on the electron gas. Due to the strong magnetic field, the SC is in the Abrikosov phase \cite{A1957} and is therefore characterized by a triangular vortex lattice with primitive lattice vectors ${\bf d}_1 = a_y {\bf e}_y$ and ${\bf d}_2 = (a_x {\bf e}_x + a_y{\bf e}_y)/2$, where $a_x= \sqrt{3} a_y$ and $a_y$ denote the lattice spacing in $x$ and $y$ direction. Due to the fixed ratio of two superconducting vortices per electronic state in the LLL, the lattice spacing $a_y = (2\pi/\sqrt{3})^{1/2} l_B$ is determined by the magnetic length. We here explicitly consider a triangular vortex lattice, however, our main results also persist in the case of a square lattice. 

We describe the two-dimensional vortex lattice by an $a_x \times a_y$ rectangular unit cell. Then, the commuting generalized translation operators \cite{B1964} are given by 
\begin{equation}
\mathcal{T}_x = e^{(-\partial_x+iy/l_B^2)a_x}, \hspace{.6cm} \mathcal{T}_y = e^{-\partial_y a_y}. \hspace{.75cm}
\end{equation}
They have eigenstates 
\begin{equation}
\mathcal{T}_x | \mathbf{k} \rangle = e^{ik_x a_x}| \mathbf{k} \rangle , \hspace{.95cm} \mathcal{T}_y | \mathbf{k} \rangle = e^{ik_y a_y}| \mathbf{k} \rangle
\end{equation}
with $| \mathbf{k} \rangle \equiv | k_x,\ k_y \rangle = \hat{c}_\mathbf{k}^\dagger |0 \rangle$ and\begin{equation}
\hat{c}_\mathbf{k}^\dagger = \sqrt{\frac{a_x}{L_x}} \sum_n e^{-ink_x a_x}\hat{a}_{k_y+nQ_y}^\dagger .
\end{equation}
Here, $k_j = 2\pi n_j/L_j$ with $n_j \in\{1, \ \dots, \ L_j/a_j\}$ and $Q_j = 2\pi/a_j$ for $j=x,y$.

The proximity of the SC to the electron gas induces an effective $p$-wave pairing
\begin{equation}
H_{\rm SC} = \int d^2 r \, \hat{\psi}^\dagger({\bf r}) \Delta({\bf r}) \left( \partial_x - i \partial_y\right) \hat{\psi}^\dagger ({\bf r}) +{\rm h.c.},
\label{eqn:SC_bare}
\end{equation}
which originates from tunneling of Cooper pairs from the SC into the electron gas. In our analysis, we consider the superconducting pairing potential
\begin{align} 
\Delta ({\bf r}) = \Delta_0 l_B \sum_{ m } (-i)^{m^2} e^{-i mQ_yy-\big(\frac{x}{l_B} - \frac{m}{2}Q_yl_B\big)^2} ,
\label{eqn:pairing_potential}
\end{align}
which is the solution of the linearized Ginzburg-Landau equation in the LLL approximation for a triangular lattice and the homogeneous magnetic field $B$ described by the vector potential ${\bf A}$ in the Landau gauge \cite{RL2010}. Under translations, we find $\mathcal{T}_x \Delta ({\bf r})\mathcal{T}_x^{-1} = e^{i2Q_yy} \Delta ({\bf r})$ and $\mathcal{T}_y \Delta ({\bf r})\mathcal{T}_y^{-1} = \Delta ({\bf r})$.

Besides the pairing described above, the proximity coupled SC will in general give rise to a single-particle potential with the symmetry of the vortex lattice, for instance due to a residual density of states within the vortex cores. In addition, we consider external potentials e.g. due to electrostatic gates. The combined potential $V({\bf r})$ with 
\begin{equation}
H_T= \int d^2 r \, \hat{\psi}^\dagger({\bf r}) V({\bf r}) \hat{\psi}({\bf r}) 
\label{eqn:V_bare}
\end{equation}
then couples the single-particle states in the LLL.

In the following, we assume that the cyclotron energy $E_c = \hbar eB/m$ is the largest energy scale, i.e. $E_c>\Delta_0 \gtrsim \overline{V}, \overline{V^2}^{1/2}, \mu$, where $\mu$ is the chemical potential relative to the LLL, and $\overline{V}, \overline{V^2}^{1/2}$ denote the average
and standard deviation of the external potential. 
Then, the low-energy electronic degrees of freedom of the combined QH/SC hybrid system are determined by the electrons in the LLL. To investigate the effect of the pairing $\Delta({\bf r})$ and the potential $V({\bf r})$ on the electrons in the LLL, we project the Hamiltonians Eqs. \eqref{eqn:SC_bare} and \eqref{eqn:V_bare} onto the LLL using Eq. \eqref{eqn:psi_LLL}. This projection yields the low-energy Hamiltonians
\begin{subequations}
\begin{align} 
H_{\rm SC} &= \sum_\mathbf{k} \Delta_\mathbf{k} \hat{c}^\dagger_\mathbf{k} \hat{c}^\dagger_{-\mathbf{k}} +{\rm h.c.},\\
H_T&= \sum_{\mathbf{k},\mathbf{k}'} t_{\mathbf{k},\mathbf{k}'} \hat{c}^\dagger_\mathbf{k} \hat{c}_{\mathbf{k}'}^{}.
\end{align}
\end{subequations}
For the superconducting pairing potential introduced in Eq. \eqref{eqn:pairing_potential}, we find the pairing matrix elements 
\begin{equation}
\Delta_\mathbf{k} = -\sqrt{2}\partial_x \Delta({\bf r})|_{{\bf r} = l_B^2 {\bf k} \times \hat{z}} , 
\label{eqn:pairing_LLL}
\end{equation}
which couple momenta $\mathbf{k}$ and $-\mathbf{k}$. This pairing potential in momentum space mainly corresponds to $\Delta({\bf r})$, which is already a solution in the LLL approximation \cite{RL2010}. The derivative in Eq.~\eqref{eqn:pairing_LLL} is in $x$-direction only since the symmetric $p$-wave type derivative in Eq.~\eqref{eqn:SC_bare} acts on LLL eigenfunctions in the Landau gauge.

Due to the derivative term in Eq. \eqref{eqn:pairing_LLL}, the extrema of $\Delta({\bf r})$ correspond to the zeroes of $\Delta_{\bf k}$, i.e. to the unpaired momenta ${\bf M}$ and ${\bf K}$ as shown in Fig. \ref{fig:BZ_2d}. The six inequivalent saddle points of $|\Delta({\bf r})|$ yield the unpaired momenta ${\bf M} \in \{(0, 0),\ (0, \pi),\ (\pi, 0),\ (\pi,\pi),\ \pm(\pi/2,\pi/2)\}$, where momenta are in units of $1/a_x$ and $1/a_y$. Similarly, the four inequivalent maxima yield the unpaired momenta ${\bf K} \in \{ \pm(\pi/2, \pi/6),\ \pm(\pi/2, 5\pi/6)\}$. We here denote the unpaired momenta by ${\bf M}$ and ${\bf K}$ since these momenta correspond to the $M$ and the $K$, $K'$ points in the hexagonal Brillouin zone of graphene \cite{NG2009}.

{\em Single particle potential.---} We consider an external potential with periods ${\bf R}_1$ and ${\bf R}_2$, and Fourier expand the potential as 
 $V(\mathbf{r}) = \sum_\mathbf{q} V_\mathbf{q} e^{i\mathbf{q}\cdot \mathbf{r}}$ with ${\bf q}\cdot {\bf R}_j = 2\pi n_j$, and $n_j\in {\mathbb Z}$. Then, the hopping matrix elements are given by
\begin{align}
t_{\mathbf{k},\mathbf{k}'} =& \sum_\mathbf{q} V_\mathbf{q} e^{-\frac{q^2l_B^2}{4}} \sum_{n_x,n_y} \delta_{{\bf k'},{\bf k}+{\bf q}-(n_xQ_x,n_yQ_y)} \nonumber \\
& \times e^{\frac{i}{2}l_B^2\{ q_x(k_y+k_y')-n_yQ_y(k_x+k_x')\}} ,
\label{eqn:hopping}
\end{align}
such that $V_{\bf q}$ couples momenta ${\bf k}$ and ${\bf k} \pm {\bf q} \ {\rm mod}\ (Q_x, Q_y)$. This potential yields a dispersion for the LLL and thereby can lift the degeneracy between the momenta ${\bf M}$ and ${\bf K}$. 

Any potential which respects the symmetry of the vortex lattice does not lift the degeneracy of states with unpaired momenta in such a way that 
an odd number of them can cross the Fermi energy: for a potential $V_{\rm latt}$ which commutes with the translation operators, i.e. in particular any potential which reflects the symmetry of the vortex lattice, using Eq.~\eqref{eqn:hopping}    we find $t_{{\bf k},{\bf k}'} = \delta_{{\bf k},{\bf k}'} t_{{\bf k},{\bf k}} \equiv \delta_{{\bf k},{\bf k}'} t_{-{\bf k},-{\bf k}}$ and the excitation spectrum $E_{\bf k} = \sqrt{(t_{{\bf k},{\bf k}} -\mu)^2 + |\Delta_{\bf k}|^2}$. Moreover, by inspection of Eq.~\eqref{eqn:hopping}  we find $t_{{\bf k}+{\bf Q}/2,{\bf k}+{\bf Q}/2}=t_{{\bf k},{\bf k}}$ for ${\bf Q} = (Q_x, Q_y)$ and all ${\bf k}$, thus that  the excitation spectrum satisfies $E_{{\bf k}+{\bf Q}/2} = E_{\bf k}$. Therefore, an even number of ${\bf M}$ and the ${\bf K}$ points is degenerate, and the system still has an even-parity ground state. Thus, we conclude that potentials $V_{\rm latt}$ are not sufficient to drive the system into a topologically nontrivial phase.

{\it Cosine modulation}.---In the following, we consider a cosine modulated electrostatic potential
\begin{equation}
V({\bf r}) = W \cos(\mathbf{q}_0\cdot \mathbf{r} + \phi_0)
\end{equation}
with wave vector $\mathbf{q}_0$ and phase shift $\phi_0$. For $\mathbf{q}_0= {\bf e}_yQ_y/2$, we find the non-vanishing matrix elements
\begin{equation}
t_{\mathbf{k},\mathbf{k}+{\bf q}_0} = W e^{-\frac{\pi\sqrt{3}}{8}} e^{-ik_xa_x/2 } \cos\left(\frac{k_xa_x}{2} + \phi_0\right)
\label{eqn:t_y}
\end{equation}
giving rise to the normal-state dispersion relation $\epsilon_{\mathbf{k},\pm} = \pm |t_{\mathbf{k},\mathbf{k}+{\bf q}_0}|$. For $\phi_0 \neq \pi/4$, the potential splits the degeneracy between the unpaired momenta at $k_y=0$ and $k_y=Q_y/2$. In particular, for $\phi_0 = 0$ the momenta $(0,0)$ and $(0,\pi)$ are coupled while all other unpaired momenta remain uncoupled. This coupling gives rise to a level splitting of two unpaired single-particle states at $ \epsilon_\pm = \pm W e^{-\frac{\pi\sqrt{3}}{8}}$. According to a theorem \cite{K2001,S2009,FB2010}, we find a topological phase transition at $\mu = \epsilon_\pm$ which separates a nontrivial phase for $0<| \mu |< \epsilon_+$ from the trivial phase for $| \mu |> \epsilon_+$. Qualitatively, these findings also persist for $\mathbf{q}_0={\bf e}_yQ_y/2n $ with $n\in {\mathbb N}$.

In Fig.~\ref{fig:cosy_pot_spec}, we show the excitation spectrum for a QH-SC hybrid system subject to a potential with cosine modulation in $y$ direction. As expected, we find a closing and reopening of the excitation gap for $\mu = 0$, where the unpaired and uncoupled states are moved through the Fermi level, and for $\mu = \epsilon_\pm$, where even and odd linear combinations of 
the coupled states $(0,0)$ and $(0,\pi)$ are moved through the Fermi level. The closing and reopening for $\mu = \epsilon_\pm$ is a signature of the topological phase transitions taking place at these points. Moreover, we find that for $0<| \mu |< \epsilon_+$ the system has a Majorana edge mode.

For a different choice of $\mathbf{q}_0= {\bf e}_xQ_x/2$, we find 
\begin{equation}
t_{\mathbf{k},\mathbf{k}+{\bf q}_0} = W e^{-\frac{\pi}{2\sqrt{3}}} \cos(k_y a_y + \phi_0) .
\end{equation}
with normal-state dispersion relation
$\epsilon_{\mathbf{k},\pm} = \pm |t_{\mathbf{k},\mathbf{k}+{\bf q}_0}|$, which for $\phi_0 \neq \pi/4$ lifts the degeneracy between states with momentum components $k_x=0$ and $k_x=Q_x/2$. However, contrary to the case Eq.~\eqref{eqn:t_y}, 
this potential does not lift the degeneracy between states with $k_y=0$ and $k_y=Q_y/2$. Thus, an even number of unpaired momenta is moved through the Fermi level for $\mu = \pm W e^{-\frac{\pi}{2\sqrt{3}}} \cos(\phi_0) $, and therefore the system remains topologically trivial. This finding is also robust against variations of $q_{0,x}$ as long as $q_{0,y} \simeq 0$.

\begin{figure}[ht]
\includegraphics[width = .45\textwidth]{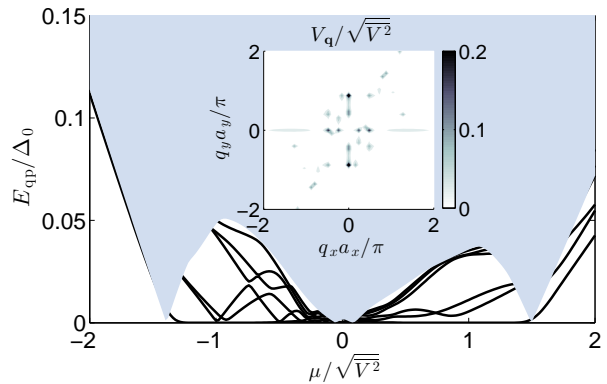}
\caption{Excitation spectrum as function of chemical potential for a potential $V({\bf r})$ with dominating Fourier coefficient for $\mathbf{q}= \pm {\bf e}_y Q_y/2 $, and Gaussian disorder with $\overline{V^2}=\Delta_0^2/6$, $L_x = 48 a_x$, and $L_y = 48 a_y$. Shaded: Bulk spectrum for periodic BC in $x$ direction. Black lines: Majorana edge mode for open BC in $x$ direction. Inset: Fourier coefficients of the potential $V({\bf r})$ as function of momentum ${\bf q}$. }
\label{fig:spectrum_disordered}
\end{figure}

In Fig. \ref{fig:spectrum_disordered}, we show the excitation spectrum for a more general external potential with dominating Fourier coefficients for ${\bf q} = \pm {\bf e}_y Q_y/2$. In addition, we add nonzero Fourier components with $q_x \neq 0$, and weak gaussian disorder. Again, we find a closing and reopening of the bulk excitation gap for $\mu = \pm |t_{\mathbf{0},\frac{Q_y}{2} {\bf e}_y}| $ and a topologically nontrivial phase with Majorana edge modes for $0<|\mu| \lesssim |t_{\mathbf{0},\frac{Q_y}{2} {\bf e}_y}| $. This demonstrates that our results described above are robust and also persist for more general potentials. 

A proximity pairing as described by Eq. \eqref{eqn:SC_bare} can be engineered by using a conventional singlet SC \cite{LS2010,OR2010} which is coupled to the QH system via a spin sensitive barrier. We model such a barrier by the Hamiltonian
\begin{align}
H_{\rm b} = & \sum_{{\bf k}, s, s'} b_{{\bf k},s}^\dagger \left\{ \epsilon_{\bf k} \delta_{s,s'} +\alpha \hat{e}_z\cdot ({\bf k} \times {\bf \hat{\sigma}}_{ss'}) +E_Z \sigma^z_{ss'} \right\} b_{{\bf k}s'}^{} \nonumber \\
& + \Delta_{\rm b} \sum_{\bf k} \left\{ b_{{\bf k}\uparrow}^\dagger b_{-{\bf k}\downarrow}^\dagger + b_{-{\bf k}\downarrow}^{} b_{{\bf k}\uparrow}^{} \right\} ,
\end{align}
where $\epsilon_{\bf k}$ denotes the bare dispersion of the film, $\alpha$ the spin orbit velocity, $E_Z$ the Zeeman field, and $\Delta_{\rm b}$ the proximity field within the barrier. To lowest order perturbation theory in the momentum conserving tunneling between the layers with amplitude $t$, and for $\epsilon_{\bf k},\Delta_{\rm b},\alpha k_F \ll E_Z$, we find
\begin{equation}
\Delta_{\rm ind}({\bf k}) = \frac{2t^2 \Delta_{\rm b} \alpha}{E_Z^3}\left(-i k_x + k_y\right), 
\end{equation}
which has the desired $p$-wave symmetry of Eq. \eqref{eqn:SC_bare}. 

An alternative scenario 
 is to start from a spin-unpolarized $\nu=2$ system and replace the pairing in Eq. \eqref{eqn:SC_bare} by the conventional spin-singlet pairing $\Delta({\bf r}) \psi_\uparrow^\dagger ({\bf r}) \psi_\downarrow^\dagger ({\bf r})$. In this case, we directly find $\Delta_\mathbf{k} = \Delta(l_B^2 {\bf k} \times \hat{z})$ which has two unpaired momenta per spin direction, corresponding to the location of the vortex cores. Then, an appropriately modulated tilted magnetic field couples the unpaired momenta and gives rise to a topological SC with pairs of Majorana edge modes. 

{\it Conclusions}.---We have developed a scenario to engineer a topological SC based on a spatially homogeneous $\nu=1$ integer QH system proximity coupled to an $s$-wave SC via a spin sensitive barrier. Due to the vortex lattice in the presence of a strong magnetic field, the hybrid system is characterized by ten unpaired generalized momenta, which have degeneracies such that always an even number of them crosses the Fermi level, implying that the hybrid system always stays in a trivial phase. Adding a periodic potential with Fourier components connecting some of the unpaired momenta, one can lift the degeneracies in such a way that the system is driven into a topologically nontrivial superconducting phase. 

{\it Acknowledgements.---}We acknowledge helpful discussions with B.I.~Halperin and financial support by DFG grants RO 2247/7-1 and RO 2247/8-1.

\end{document}